\documentclass[aps,prx,twocolumn,superscriptaddress]{revtex4-1}

\usepackage{verbatim}
\usepackage{amsmath}
\usepackage[utf8]{inputenc}
\usepackage[pdftex]{graphicx}
\usepackage[thinspace]{SIunits}
\usepackage[colorlinks=true,urlcolor=blue,linkcolor=blue,citecolor=blue]{hyperref}
\usepackage{color}
\usepackage{placeins}
\usepackage[T1]{fontenc}
\usepackage{color}
\definecolor{red}{rgb}{1.0,0.0,0.0}
\definecolor{blue}{rgb}{0.0,0.0,1}
\definecolor{green}{rgb}{0.29, 0.33, 0.13}

\begin{document}

\title{Coexistence of weak and strong coupling with a quantum dot in a photonic molecule}

\author{S.~Lichtmannecker}
\affiliation{
Walter Schottky Institut and Physik Department, 
Technische Universit\"at M\"unchen, Am Coulombwall 4, 85748 Garching, Germany}
\author{M.~Kaniber}%\email{michael.kaniber@wsi.tum.de}
\affiliation{
Walter Schottky Institut and Physik Department, 
Technische Universit\"at M\"unchen, Am Coulombwall 4, 85748 Garching, Germany}
\author{S.~Echeverri-Arteaga} \affiliation{Departamento de F\'isica, Universidad Nacional de Colombia, 111321, Bogot\'a, Colombia}
\author{I.~C.~Andrade} \affiliation{Universidad de Antioquia,
  Instituto de F\'isica, Medell\'in, AA 1226 Medell\'in, Colombia}

\author{J.~Ruiz-Rivas}
\affiliation{
Departament d’\`{O}ptica, Universitat de Val\`{e}ncia, Dr. Moliner 50, 46100 Burjassot, Spain}
\author{T.~Reichert}
\affiliation{	
Walter Schottky Institut and Physik Department, 
Technische Universit\"at M\"unchen, Am Coulombwall 4, 85748 Garching, Germany}
\author{M.~Becker}
\author{M.~Blauth}
\author{G.~Reithmaier}
\affiliation{
Walter Schottky Institut and Physik Department, 
Technische Universit\"at M\"unchen, Am Coulombwall 4, 85748 Garching, Germany}
\author{P.~L.~Ardelt}
\affiliation{
Walter Schottky Institut and Physik Department, 
Technische Universit\"at M\"unchen, Am Coulombwall 4, 85748 Garching, Germany}
\author{M.~Bichler}
\affiliation{
Walter Schottky Institut and Physik Department, 
Technische Universit\"at M\"unchen, Am Coulombwall 4, 85748 Garching, Germany}

\author{E.~A.~G\'omez} \affiliation{Programa de F\'isica, Universidad del Quind\'io, 630004, Armenia, Colombia}
\author{H.~Vinck-Posada} \affiliation{Departamento de F\'isica, Universidad Nacional de Colombia, 111321, Bogot\'a, Colombia}

\author{E.~del Valle}
\affiliation{
Departamento de F\'{i}sica Te\'{o}rica de la Materia Condensada, 
Universidad Aut\'{o}noma de Madrid, E-28049 Madrid, Spain}

\author{J.~J.~Finley}\email{finley@wsi.tum.de}
\affiliation{
Walter Schottky Institut and Physik Department, 
Technische Universit\"at M\"unchen, Am Coulombwall 4, 85748 Garching, Germany}

\date{\today}

%###############################################################################
%									ABASTRACT
%###############################################################################
\begin{abstract}
  We study the emission from a molecular photonic cavity formed by two
  proximal photonic crystal defect cavities containing a small number
  (<3) of In(Ga)As quantum dots. Under strong excitation we observe
  photoluminescence from the bonding and antibonding modes in
  excellent agreement with expectations from numerical
  simulations. Power dependent measurements reveal an unexpected
  peak, emerging at an energy between the bonding and antibonding modes
  of the molecule. Temperature dependent measurements show that this
  unexpected feature is photonic in origin.  Time-resolved
  measurements show the emergent peak exhibits a lifetime
  $\tau_M=\unit{0.75 \pm 0.1}{\nano\second}$, similar to both bonding
  and antibonding coupled modes. Comparison of experimental results
  with theoretical expectations reveal that this new feature arises
  from a coexistence of weak- and strong-coupling, due to the molecule
  emitting in an environment whose configuration permits or, on the
  contrary, impedes its strong-coupling.  This scenario is reproduced
  theoretically for our particular geometry with a master equation
  reduced to the key ingredients of its dynamics. Excellent
  qualitative agreement is obtained between experiment and theory,
  showing how solid-state cavity QED can reveal new regimes of
  light-matter interaction.
\end{abstract}
%
%###############################################################################
%
\maketitle

\section{Introduction}

Cavity QED (cQED) in solid-state systems \cite{arakawa15a} is rapidly
developing into a field of its own following the Nobel prize winning
precedent set by atoms in microwave cavities~\cite{haroche13a}.
Unlike their atomic counterparts, solid state systems provide great
flexibility to engineer ad hoc structures in complex geometries \cite{kuruma16a}. Among
the possible architectures, photonic crystal (PhC) nanostructures triplet peak structure
provide the flexibility to probe cQED phenomena in non-standard
configurations~\cite{lodahl2015interfacing}.  Due to their planar
geometry, they provide a promising platform for future integrated
quantum photonic devices~\cite{o2009photonic}. High quality (Q)
factors combined with the ultra-small mode volumes of PhC cavities
allows cQED to be studied in the few photon limit
\cite{soljavcic2004enhancement,englund2007controlling,fushman2008controlled,englund2010deterministic,volz2012ultrafast,englund2012ultrafast,choi17a}.
Most of the cQED experiments performed to date using PhCs have been
performed using a single cavity. In this work, by coupling two
proximal nano-resonators to form a photonic molecule (PM)
\cite{bayer1998optical,atlasov2008wavelength,dousse2010ultrabright,chalcraft2011mode,atlasov2011large,majumdar2012cavity},
we open the way to explore new degrees of freedom with potential for
entirely new functionalities. For example the energy splitting of the
PM modes can be tuned via geometric parameters during fabrication or
tuning using photochromic materials or nanoelectromechanical systems
\cite{cai2013controlled,haddadi14a,kapf2015dynamic,du16a}. This allows the simultaneous
enhancement of two different transitions and establishing of coupling
between two quantum emitters separated by distances comparable to the
optical wavelength
\cite{dousse2010quantum,dousse2010ultrabright}. Recent theoretical
proposals taking advantage of coupled resonators suggest new
applications, such as the generation of optimized Gaussian amplitude
squeezing with very small Kerr nonlinearities~\cite{liew2010single},
the generation of bound photon-atom states~\cite{longo2010few} or the
full optical coherent control of vacuum Rabi
oscillations~\cite{bose2014all}. Photonic crystal molecules are also
of great interest for solid-state implementations of photonic quantum
simulators~\cite{hartmann2006strongly,greentree2006quantum,houck2012chip}.
However, to date, only a
handful of experiments have been performed using PMs,
exploring non-linear effects such as sum frequency
generation~\cite{rivoire2010sum,rivoire2011multiply} or parametric
oscillation~\cite{armstrong1962interactions,diederichs2006parametric},
despite the early demonstration of the up-conversion 
excitation in bulk GaAs~\cite{kammerer2001photoluminescence}, 
and enhanced efficiencies using planar
microcavities~\cite{diederichs2006parametric,xu2014giant}.

Here, we investigate the linear and non-linear properties of an
individual PM formed by two coupled PhC cavities doped with self
assembled quantum dots (QDs). By performing photoluminescence (PL) and
PL-excitation (PLE) spectroscopy we provide clear evidence for the
photonic coupling of the two cavities. In power dependent
PL-measurements we observe bonding- (B) and antibonding- (AB) like
modes of the PM at energies that are in excellent quantitative
agreement with finite-difference time-domain (FDTD)
simulations. Surprisingly, we observe an additional unexpected peak
(W) that emerges precisely between B and AB when the system is
subjected to strong excitation. Time-integrated PL measurements
performed as a function of the lattice temperature and time-resolved
spectroscopy reveal that this additional unexpected peak is primarily
\textit{photonic} in origin.  We explain this unexpected feature as a
zero-dimensional counterpart of phonon-sidebands where an optical
transition occurs in a lattice environment which is altered by the
emission itself, making it dependent on whether the emitter is in its
ground or excited state. Here, in addition to substituting the phonon
bath by a two-level system, the emission itself is from a
strongly-coupled system which features a Rabi doublet instead of a
single line. This results in a peculiar phenomenology where an
anomalous peak seem to grow in between a conventional Rabi doublet,
that our interpretation shows results from a coexistence of weak and
strong-coupling, as an extreme case where the molecule finds itself in
an environment that either exposes or shields it from an additional
decay channel which results in spoiling or preserving its coherent
Rabi dynamics. A quantum-optical model that couples a QD to the PM
through phonon-mediated transitions captures this phenomenon and
provides a fundamental picture of this otherwise peculiar
mechanism. Our result shows that the highly complex configurations one
can engineer in the solid state provide interesting variations on the
basic themes of light-matter interactions.

%##################### FABRICATION EXPERIMENTAL SETUP #########################
\section{Fabrication and Experiment}

\begin{figure}
  \includegraphics{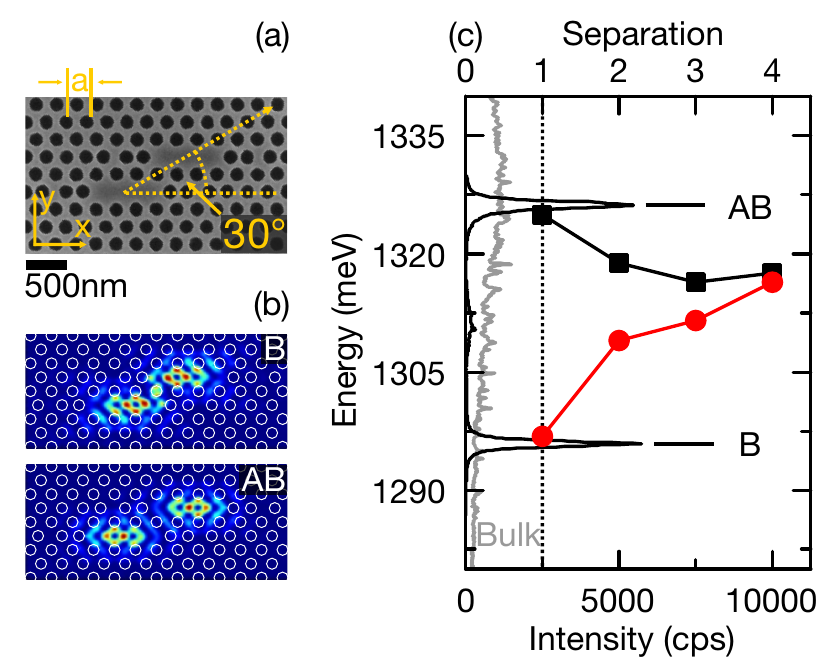}
  \caption{ \label{figure1_SEM}
    (a) SEM image of the PM formed by two L3 cavities.
    (b) $|E|^2$ extracted from FDTD simulations for the B (upper panel) and AB
    (lower panel) mode. The white circles indicate the air holes forming the PhC.
    (c) PL spectrum of the investigated PM (black curve), and QD
    emission from the unpatterned region of the sample (gray
    curve). The solid red dots and black squares indicate the
    simulated energetic position of the AB and B modes as a function
    of cavity separation.}
\end{figure}

The sample was grown using molecular beam epitaxy on a
$\unit{350}{\micro\meter}$ thick [100] orientated GaAs wafer. After
depositing a $\unit{300}{\nano\meter}$ thick GaAs buffer layer we grew
a $\unit{800}{\nano\metre}$ thick sacrificial layer of
Al$_{0.8}$Ga$_{0.2}$As, followed by a $\unit{150}{\nano\metre}$ thick
nominally undoped GaAs waveguide containing a single layer of
In$_{0.5}$Ga$_{0.5}$As QDs at its midpoint.  The growth conditions
used for the QD layer produce dots with an areal density
$\rho_D \sim $ $\unit{5}{\micro\meter^{-2}}$, emitting over the energy
range $E_{QD}=\unit{1260-1400}{\milli\electronvolt}$. After growth, a
hexagonal lattice of air holes with a lattice constant of
$a=\unit{260}{\nano\meter}$ was defined in a ZEP $520$-A soft mask and
deeply etched using a SiCl$_4$ based inductively coupled plasma to
form a two-dimensional PhC. The resulting PM is formed by two L3 cavities
\cite{akahane2003design} with their edges separated by a single period
of the PhCl lattice as shown by the scanning electron
microscopy image in figure \ref{figure1_SEM}(a). In a final step, the
AlGaAs layer was selectively removed with hydrofluoric acid to
establish a free standing membrane.

After fabrication and characterization the sample was cooled to a
lattice temperature $T=\unit{13}{\kelvin}$ in a He flow-cryostat for
optical study. Thereby, we used a $100\times$ microscope objective
with a numerical aperture $\text{NA}=0.8$ in a confocal geometry
provided by coupling the emitted signal into a single mode fiber to
spatially detect emission from a region of interest with a size of
$\unit{1}{\micro\meter}\times\unit{1}{\micro\meter}$. The sample was
optically excited using a pulsed laser with a repetition frequency of
$\unit{80}{\mega\hertz}$. Hereby, either a non-resonant diode laser at
$\unit{1580}{\milli\electronvolt}$ ($\unit{80}{\pico\second}$ pulse
duration), or a Ti:sapphire laser with an emission energy tuned to
$\unit{1312}{\milli\electronvolt}$ and a pulse duration of
$\unit{10}{\pico\second}$ was used. For measurements using cavity mode
resonant excitation \cite{nomura2006localized,kaniber2009efficient} we
used a tunable continuous wave single frequency laser with a bandwidth
of $\unit{100}{\kilo\hertz}$. The collected emission from the sample
was spectrally dispersed using an imaging monochromator with a focal
length of $\unit{0.5}{\meter}$ and detected with a liquid nitrogen
cooled CCD camera. For time-resolved measurements a Si-avalanche
photodiode was used, providing a temporal resolution of
$\unit{\sim 350}{\pico\second}$ without deconvolution.

For the system of two coupled L3 cavities, we expect the formation of
bonding (B) and anti-bonding (AB) modes with even and odd symmetry, respectively
\cite{chalcraft2011mode}. FDTD simulations \cite{Lumerical2015} for
cavities having different separations and relative orientations
revealed that the $\unit{30}{\degree}$ configuration between the
L3-caviy axis and a line connecting the cavity centers provides the
strongest coupling for a given nominal separation
between cavity centers \cite{chalcraft2011mode}. Simulations using
geometrical parameters extracted from the scanning electron microscopy
image shown in figure \ref{figure1_SEM} (a) yield the electric
field distribution of the B and AB modes and their relative energies
presented in figure \ref{figure1_SEM} (b).  The energy splitting
between these two modes is plotted in figure \ref{figure1_SEM} (c)
versus the cavity-cavity separation. For a cavity separation of one
row of air holes we expect an energy splitting of
$\Delta E_{sp}^{th} = \unit{28}{\milli\electronvolt}$. For comparison
we plot in figure \ref{figure1_SEM} (c) the PL emission recorded from
the investigated PM (black curve) using strong excitation and the QD
emission from an unpatterned region of the sample as a reference (gray
curve). We clearly observe emission from the B and AB modes with an
energy splitting of
$E_{sp}^{exp}=\unit{30 \pm 0.1}{\milli\electronvolt}$, in fair
quantitative agreement with our simulations
\cite{atlasov2011large,chalcraft2011mode,majumdar2012cavity}. The
Q-factors of the B and AB modes were measured to be $\sim 1700$ and
$\sim 1400$, respectively.

%
%############################## Figure 1 SEM #################################
%##############################################################################
%
%
%######################### RESULTS AND DISCUSSIONS ############################

\section{Results and Discussions}

\begin{figure*}
\includegraphics{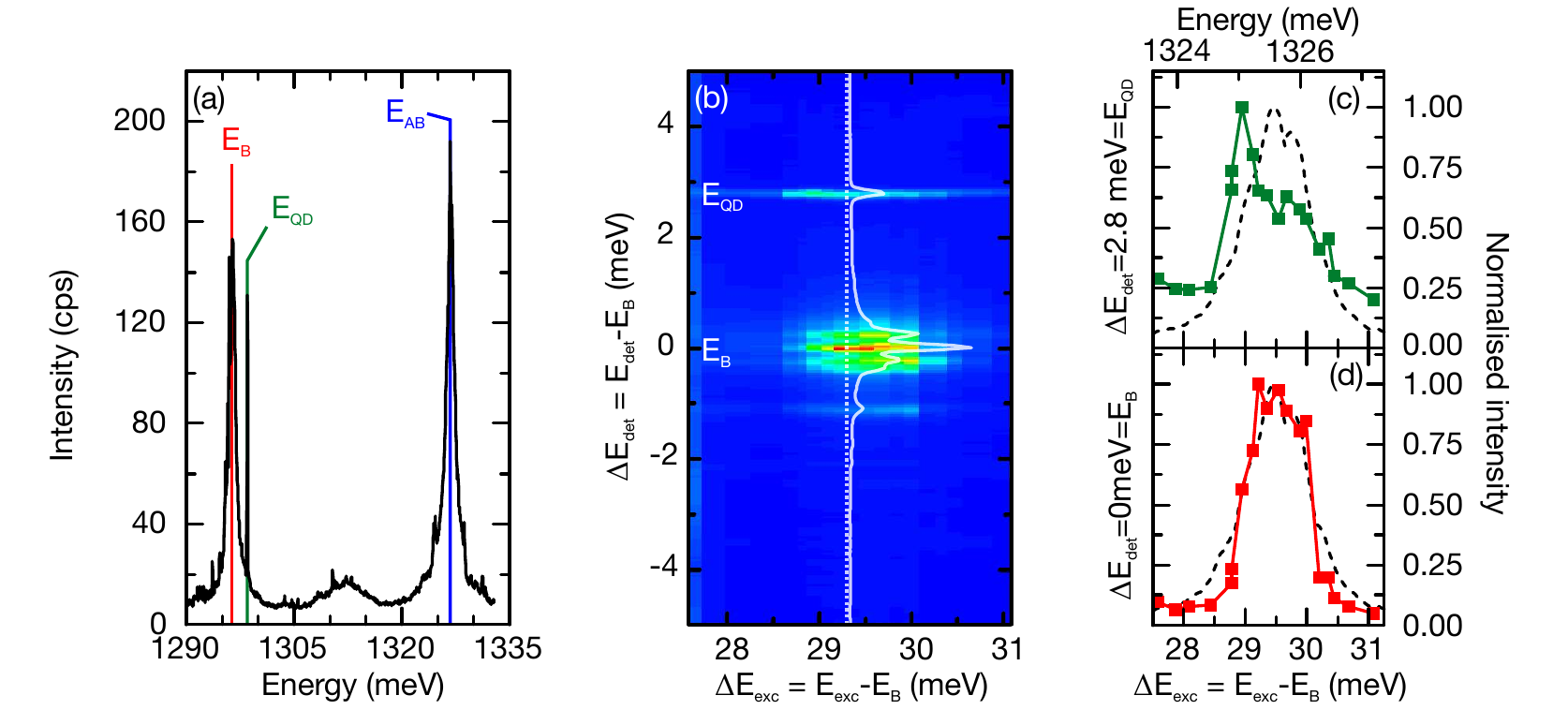}
\caption{\label{figure2_mode_PLE} 
(a) PL emission spectrum of the PM under pulsed above band gap excitation. On
the high energy side of the B mode the emission line of a QD, depicted in green,
is observed with a detuning of $\Delta E_{QD}=\unit{+2.8}{\milli\electronvolt}$.
The according energies are labeled with B, QD, and AB.
(b) Emission intensity as a function of excitation laser energy, whilst tuning
the laser across the antibonding resonance. The white line shows an emission
spectrum for the excitation energy marked by the white dotted line.
(c) Emitted intensity at the QD energy at  $\Delta
E_{det}=\unit{+2.8}{\milli\electronvolt}$ as a function of laser detuning. The
dashed line shows PL emission of the AB mode subjected to non-resonant
excitation for comparison.
(d) Emitted intensity of the bonding mode at
$\Delta E_{det}=\unit{0}{\milli\electronvolt}$. The dashed line shows
PL emission of the AB subjected to non-resonant excitation mode for
comparison.  }
\end{figure*}

In figure \ref{figure2_mode_PLE} (a) we present a typical \micro-PL
spectrum recorded from the PM under pulsed non-resonant
excitation. Again, we observe the emission of the B and AB modes as
well as a sharp emission line attributed to a single QD at
$\unit{1298.5}{\milli\electronvolt}$, depicted in green, with a
detuning of
$\Delta E_{QD} = E_{QD}-E_B = \unit{+2.8}{\milli\electronvolt}$
relative to the energy of the B mode ($E_B$). The respective relative
energies of the B mode $E_B$ (red), the QD $E_{QD}$ (green) and the AB
mode $E_{AB}$ (blue) are labeled on the figure.
To demonstrate that spatially delocalized molecular-like modes are
formed in the PM we excited the system at resonance at the AB mode
energy. This allows us to directly pump the cavity mode and excite QDs
that are located at positions close to the electric field antinodes
within either of the two
cavities~\cite{nomura2006localized,kaniber2009efficient}. To do this,
we tuned a single frequency laser across the emission energy of the AB
mode from $\unit{1327.2}{\milli\electronvolt}$ to
$\unit{1324.0}{\milli\electronvolt}$ in steps of
$\unit{200}{\micro\electronvolt}$. Simultaneously, we detected
emission spectra in the spectral vicinity of the B mode.
Figure~\ref{figure2_mode_PLE}(b) shows the color-coded emission
intensity as a function of the detection energy relative to the B
mode, $\Delta E_{det} = E_{det} - E_B$, and the excitation laser
energy $\Delta E_{exc} = E_{exc} - E_B$, for a pump power density of
$\unit{94.5}{\watt\per\centi\meter\squared}$. We observe two clear
maxima at $\Delta E_{det} = \unit{2.8}{\milli\electronvolt}$ and
$\Delta E_{det} = \unit{0}{\milli\electronvolt}$ when resonantly
exciting via the AB mode
($\Delta E_{exc} = \unit{29.3}{\milli\electronvolt}$) attributed to
the QD and the B mode, respectively.  The white line shows an emission
spectrum for the resonance condition
$\Delta E_{exc}=\unit{29.3}{\milli\electronvolt}$ indicating that the
QD and the bonding mode are simultaneously excited via the
higher-energy AB mode. In figures \ref{figure2_mode_PLE}(c) and (d),
we compare horizontal cross-sections through the QD and B mode
emission at $\Delta E_{det}=\unit{2.8}{\milli\electronvolt}$ and
$\Delta E_{det}=\unit{0}{\milli\electronvolt}$, respectively. For both
detection energies we simultaneously observe a clear maximum when
resonantly exciting via AB. The dashed black lines show the PL
spectrum of the AB mode for comparison. The observation of a shared
absorption resonance for both the QD and B mode confirms that the two
cavities are indeed coupled and that the QD is spatially coupled to
one of the two cavities forming the PM.

\begin{figure}
\includegraphics{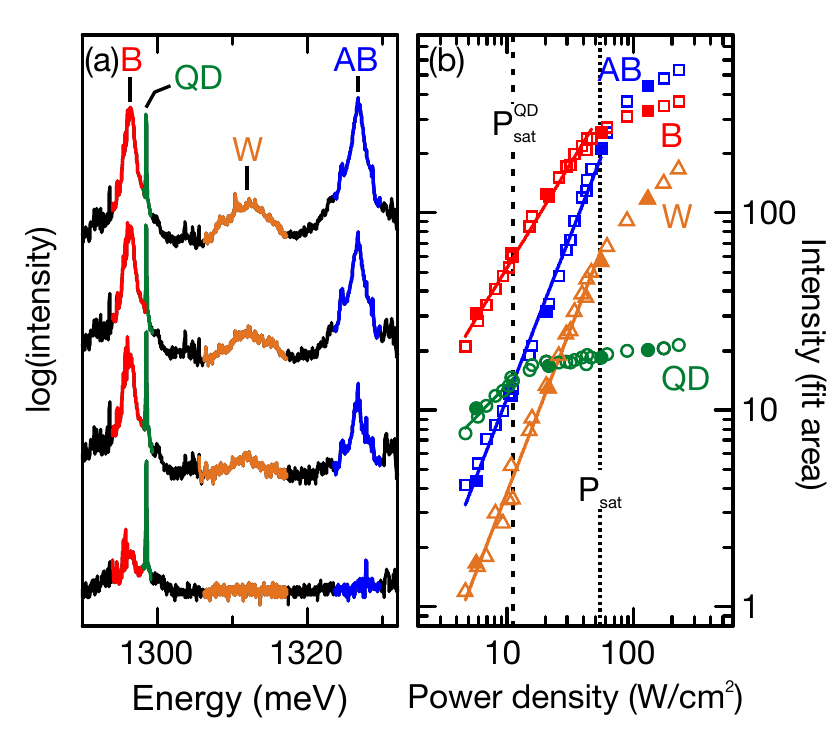} 
\caption{ \label{figure3_non_res_PL}
(a) PL intensity of the investigated PM plotted on an logarithmic scale when subjet to pulsed non-resonant excitation.
(b) Extracted integrated intensities of B, AB, QD and W as a function of the
pump power density, shown in double logarithmic representation. The excitation power densities for the plotted
spectra in (a) are highlighted with filled symbols.
the black dashed and dotted lines show the saturation power density of the QD
$P_{sat}^{QD}$ and the common saturation power $P_{sat}$ of the B, AB and W
peak, respectively. The solid lines represent power-law fits to the data points.
}
\end{figure}

After confirming the coupled character of the two cavities forming the
PM, we present detailed investigations of the linear and non-linear
optical properties of the PM coupled to the
QD. Figure~\ref{figure3_non_res_PL}(a) shows typical emission spectra
from the PM subject to non-resonant pulsed excitation at
$\unit{1580}{\milli\electronvolt}$ as the excitation power density is
increased from $\unit{\sim 2}{\watt\per\centi\meter\squared}$ to
$\unit{>200}{\watt\per\centi\meter\squared}$.  As discussed already in
the previous paragraph, we observe pronounced emission from B, AB and
the QD. More strikingly, an additional emission feature, labeled W in
figure~\ref{figure3_non_res_PL}(a), emerges for elevated excitation
power densities. The unexpected W emission is energetically centered
precisely between B and AB at
$E_M = \unit{1311}{\milli\electronvolt} \approx (E_{AB} + E_B)/{2}$.
In figure \ref{figure3_non_res_PL}(b) we present the integrated peak
intensities of B, AB, QD and W as a function of the excitation power
density, plotted on a double logarithmic representation. The filled
symbols label the excitation power densities selected for the spectra
plotted in figure \ref{figure3_non_res_PL}(a). The QD transition
increases linearly with excitation power density, followed by
saturation of the emission for excitation power densities above
$P_{sat}^{QD}=\unit{13\pm2}{\watt\per\centi\meter\squared}$, as
indicated by the dashed line in figure
\ref{figure3_non_res_PL}(b). The B mode also increases linearly with
an exponent of $1.07 \pm 0.02$, due to non-resonant feeding via QD
ground states
\cite{laucht2010temporal,hohenester2010cavity,winger2009explanation}.
In contrast, we observe for the AB mode a clear superlinear increase
in intensity of $1.65 \pm 0.03$, most likely arising from its
proximity to excited QD states.
This attribution is supported by
time-resolved measurements discussed in detail in figure
\ref{figure5_TR_spec}(a). Both A and AB modes saturate at comparable
power densities of
$P_{sat}=\unit{54\pm4}{\watt\per\centi\meter\squared}$ highlighted
with the dotted line in figure \ref{figure3_non_res_PL}(b). For the
unexpected W peak, we observe a super-linear exponent of
$1.68 \pm 0.03$, despite being at lower energy than expected for the
QD excited states.
Moreover, the W peak exhibits a similar saturation power density
$P_{sat}=\unit{54\pm4}{\watt\per\centi\meter\squared}$ as the B and AB
modes. This excitation power is $4.3\times$ higher than the saturation
power $P_{sat}^{QD}$ observed for the QD. % We conclude that the PM
%
%
%########################### Figure 2 mode PLE ################################
%##############################################################################
%

%
%
%########################## Figure 3 non res PL ###############################
%##############################################################################
%

\begin{figure}
\includegraphics{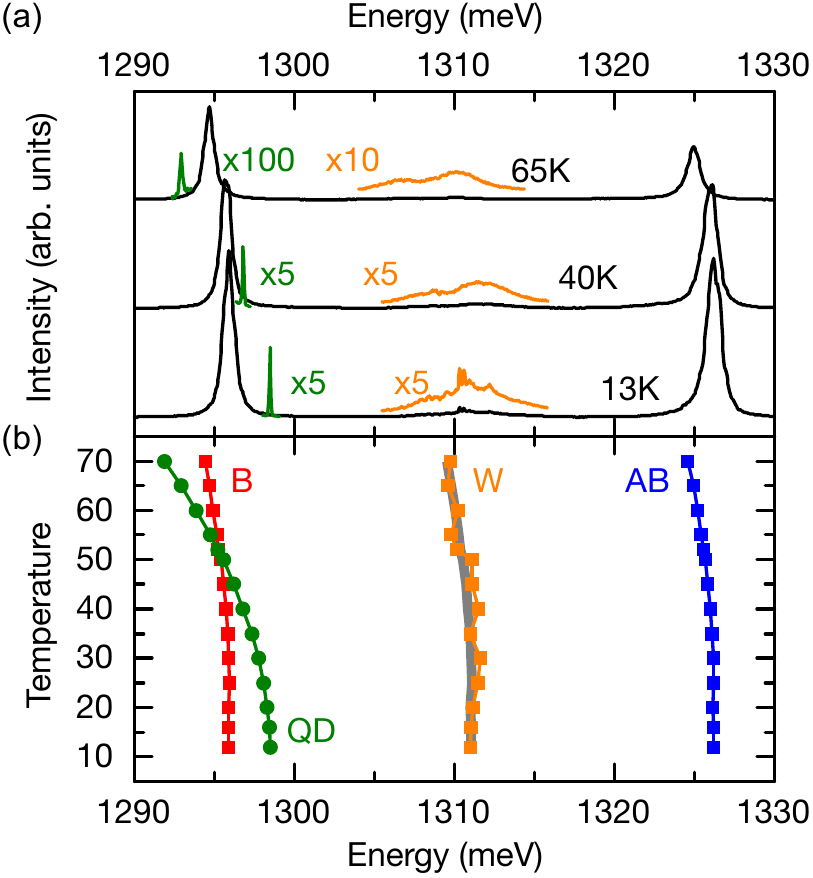}
\caption{\label{figure4_T_tuning}
(a) PL emission of the PM for selected temperatures with an excitation power
density of $\unit{99}{\watt\per\centi\meter\squared}$ (black curves), magnified
W peak (orange curves) and the QD emission under an excitation power of
$\unit{3.2}{\watt\per\centi\meter\squared}$ (green curves).
(b) Peak energies of the B mode (red), the QD (green), the W (orange)
and the AB mode (blue) as a function of crystal temperature. Grey
shaded region marks the calculated center between B and AB.  }
\end{figure}

\begin{figure}
\includegraphics{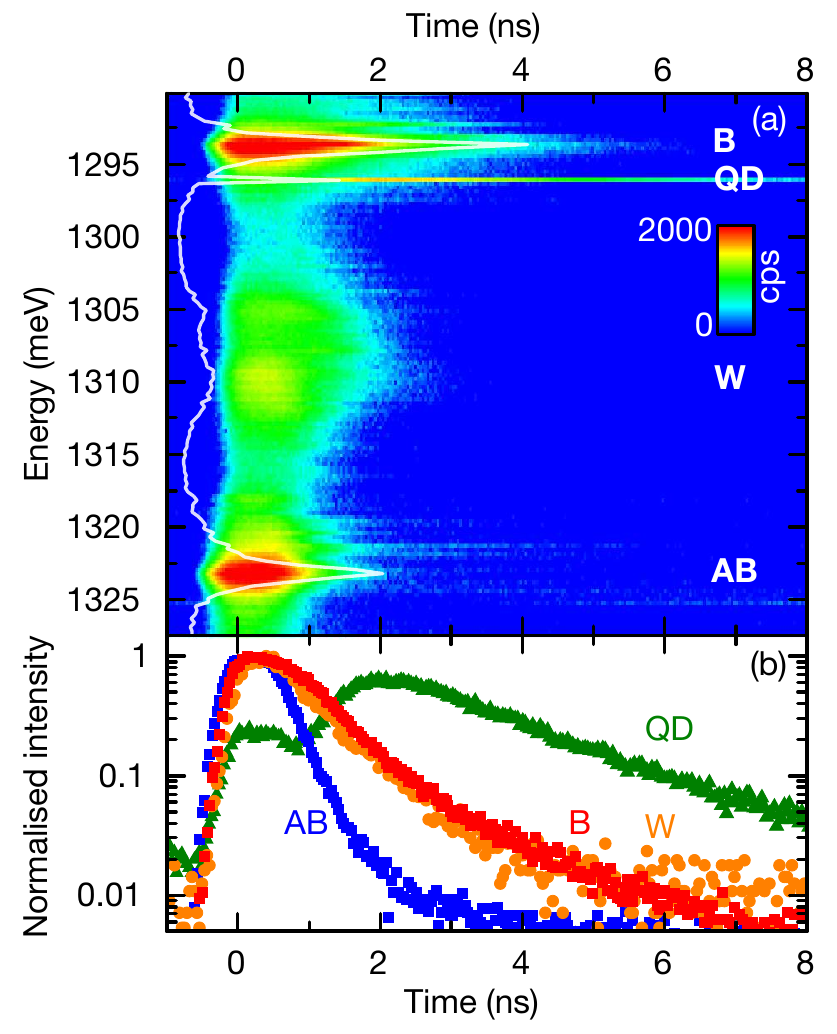}
\caption{ \label{figure5_TR_spec}
(a) Energy and time resolved \micro-PL measurements of the PM. The color scale
indicates the intensity of the recorded PL signal increasing from blue to red.
The white line represents the time-integrated intensity over all collected
times.
(b) Selected decay transients for AB, B, QD and W under non-resonant excitation.
}
\end{figure}

In order to shed light on the origin of the W peak, we present in the
following temperature dependent PL measurements which enable us to
clearly distinguish between the QD-excitonic or photonic character of
the individual emission features. For the cavity modes, we expect a
weak and approximately linear shift with increasing temperature, due
to the change in the refractive index with increasing temperature
\cite{marple1964refractive}. In contrast, the QD is expected to
exhibit a significantly stronger shift determined by a Varshni type
relation $E_{gap}(T)=E_{gap}(T=0)-(\alpha T^2)/(T+\beta)$, where
$\alpha$ and $\beta$ are dependent on the material.
% 	???3 to end note?
For GaAs $\alpha=8.871 \times 10^{-4}$, $\beta = 572$
\cite{varshni1967temperature} and
$E_{gap}(T=0)= \unit{1521.6}{\milli\electronvolt}$
\cite{sturge1962optical,thurmond1975standard} and for InAs
$\alpha=3.158 \times 10^{-4}$, $\beta = 93$ and
$E_{gap}(T=0)= \unit{426}{\milli\electronvolt}$
\cite{varshni1967temperature,dixon1961optical}.
In figure \ref{figure4_T_tuning}(a), we present PL spectra recorded
from the cavity modes (black curves) and a magnified region around W
(orange curves), as well as the QD emission (green curves) for three
selected crystal temperatures, $\unit{13}{\kelvin}$,
$\unit{40}{\kelvin}$ and $\unit{65}{\kelvin}$ and two excitation power
densities; $\unit{99}{\watt\per\centi\meter\squared}$ (black and
orange curves) and $\unit{3.2}{\watt\per\centi\meter\squared}$ (green
curves), respectively.  For both cavity modes and the QD we observe
clear shifts to lower energy with increasing lattice
temperature. However, the QD exhibits a higher shift rate. In figure
\ref{figure4_T_tuning}(b) we present the extracted peak positions of
the different emission lines whilst tuning the sample temperature from
$\unit{13crystal temperatures}{\kelvin}$ to $\unit{70}{\kelvin}$ in steps of
$\unit{5}{\kelvin}$.  For the QD, we obtain a clear non-linear shift
of the emission with temperature, yielding an average shift rate of
$\unit{113.6}{\micro\electronvolt\per\kelvin}$. As expected, the
average shift rates for the cavities modes A and AB are
$\unit{25.4}{\micro\electronvolt\per\kelvin}$ and
$\unit{27.6}{\micro\electronvolt\per\kelvin}$, respectively, and thus
a factor ${}\times 4.3$ smaller as compared to the QD. The pronounced
difference in shift rates for QD and B leads to a clear resonance for
a temperature of $T=\unit{52}{\kelvin}$. We observe that the W peak
(orange) shows an average shift of
$\unit{21.1}{\micro\electronvolt\per\kelvin}$, similar to B and AB and
stays centered between both modes over the whole temperature range
studied, as supported by the calculated center energy $E_M$ shown in
gray in figure \ref{figure4_T_tuning} (b). This demonstrates that the
observed peak W is predominantly photonic-like and most likely does
not arise from excitonic QD states.

%############################## Figure 4 T tuning #############################
%##############################################################################
%

Before identifying the nature of the unexpected W peak by comparing
our results with a theoretical model, we continue to explore the decay
dynamics of the coupled QD-PM using time-resolved
spectroscopy. Measurements were performed as a function of emission
energy
($\unit{1290}{\milli\electronvolt} < E <
\unit{1327}{\milli\electronvolt}$)
subject to non-resonant excitation at
$\unit{1579}{\milli\electronvolt}$ and a repetition frequency of
$\unit{80}{\mega\hertz}$. We used a spectrometer as tunable
$\Delta = \unit{270}{\micro\electronvolt}$ bandpass filter and
recorded time transients using a time-correlated single-photon
counting module
\cite{laucht2010temporal}. Figure~\ref{figure5_TR_spec}(a) shows the
complete detection energy- and time-resolved PL map. The white curve
represents the time-integrated signal over all recorded times and
resembles the typical PL spectra recorded with a CCD camera. We
clearly observe the B and AB mode, as well as the QD that shows a
$\sim 3$ times slower decay and, thus, is still visible in the time
transient when the signal of the cavity modes have completely decayed.
% Also in this representation of the data the delayed maximal signal of the M peak is visible.
Both the B and AB modes exhibit fast decays, from which we extract
exciton lifetimes of $\tau_{B}=\unit{0.76\pm 0.1}{\nano\second}$ and
$\tau_{AB}=\unit{0.35\pm0.1}{\nano\second}$, respectively. The shorter
lifetime for the AB cavity mode is most likely caused by the spectral
overlap with excited QD states \cite{laucht2010temporal}. For the QD
emission we observe a step-like increase in intensity as shown in
figure \ref{figure5_TR_spec}(b), accompanied by a delayed onset of the
luminescence decay. Moreover, a clear anti-correlation between the AB
and the QD signal is observed; the AB mode has fully decayed prior to
the QD decay. Both observations strongly suggest that the AB mode is
predominantly fed from energetically higher excited multi-exciton
states \cite{laucht2011nonresonant}. In figure
\ref{figure5_TR_spec}(b), we present selected decay transients of the
B (red) and AB (blue) modes, the QD (green), as well as the W (orange)
peak, as labeled in figure \ref{figure5_TR_spec}(a). The exciton
lifetime from the delayed QD decay yields
$\tau_{QD}=\unit{2 \pm0.1}{\nano\second}$. The W peak, represented by
the orange symbols in figure \ref{figure5_TR_spec}(b), shows a
lifetime similar to the B and AB modes with
$\tau_{M}=\unit{0.75\pm 0.1}{\nano\second}$, supporting again our
conclusion of the photonic-like origin of the W emission.

%
%###################### Figure 5 time resolved spectrum #######################
%##############################################################################
%

\begin{figure}
\includegraphics{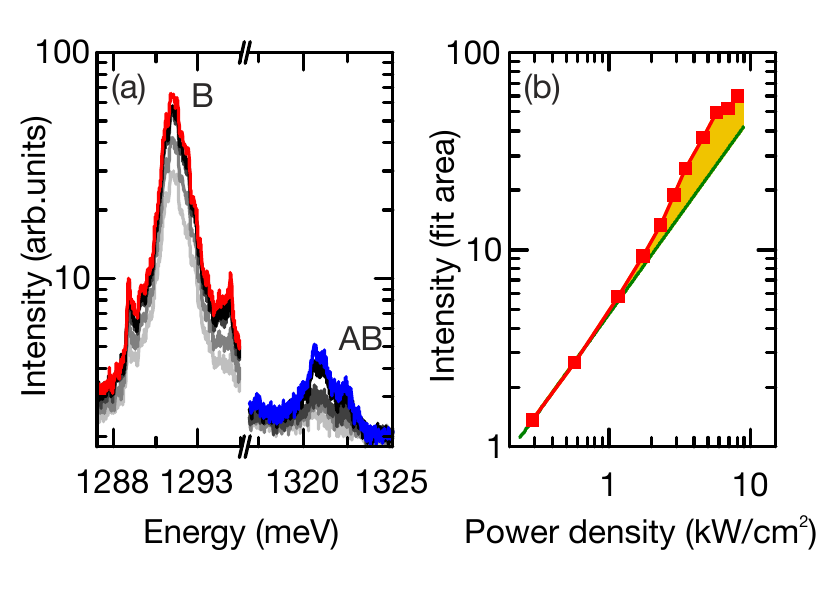}
\caption{ \label{figure7_2P_action}
(a) \micro-PL spectra detected from the PM under excitation with
$\unit{10}{\pico\second}$ pulses resonant to the W peak at
$\unit{1311}{\milli\electronvolt}$ for different optical pump powers.
(b) PL intensity of AB as a function of excitation power density in double
logarithmic representation depicted in red. The green line shows a linear
power-law fitted to the first data points. The yellow area highlights the non linear
fraction of the emission.
}
\end{figure}

Finally, we performed CW resonant excitation of the W mode and power
densities increasing from $\unit{290}{\watt\per\centi\meter\squared}$
up to $\unit{8}{\kilo\watt\per\centi\meter\squared}$. The result,
shown in figure \ref{figure7_2P_action}(a), clearly shows emission
from both B and AB.  Although the excitation laser is tuned to lower
energy than the AB mode, we detect significant emission from the
higher energy AB mode (blue). The B mode can be directly excited via
linear absorption of the excited states of the QD, since the laser is
tuned to higher energy than the emission. In figure
\ref{figure7_2P_action}(b) we plot the PL intensity of AB as a
function of the excitation power density. For low excitation power
densities ($P < \unit{1}{\kilo\watt\per\centi\meter\squared}$) we
observe a linear dependence of the emission from AB with an exponent of
$1.04 \pm 0.04$. However, for higher excitation power densities
($P > \unit{2.5}{\kilo\watt\per\centi\meter\squared}$) the emission
becomes super-linear, with the yellow shaded region highlighting the
difference between the linear and super-linear behavior. This
indicates that the W mode is directly coupled to the B and AB modes,
but there is a-priori no mechanism to account for this coupling. In
the following, we will discuss how this arises from a coexistence of
weak and strong coupling of the molecule.

\section{Identification of the anomalous peak}

The observation of a triplet peak structure in solid-state in
strong-coupling experiments \cite{hennessy07a,winger08a,ota09b,ota18a} has been a recurrent conundrum for
theorists~\cite{gonzaleztudela10b,yamaguchi09a,yamaguchi12a}.
Like in some other cases where a spectral triplet was observed when
only a Rabi doublet was expected, our explanation
relies on a mixture of weak and strong coupling. But instead of a mere
incoherent superposition of the two regimes, that would be observed
independently in separate time windows, our system involves an
inextricable coexistence where both the weak and strong coupling occur
simultaneously or at least during the smallest timescale of the
system dynamics.
\begin{figure}[t]
  \centering
  \includegraphics[width=\linewidth]{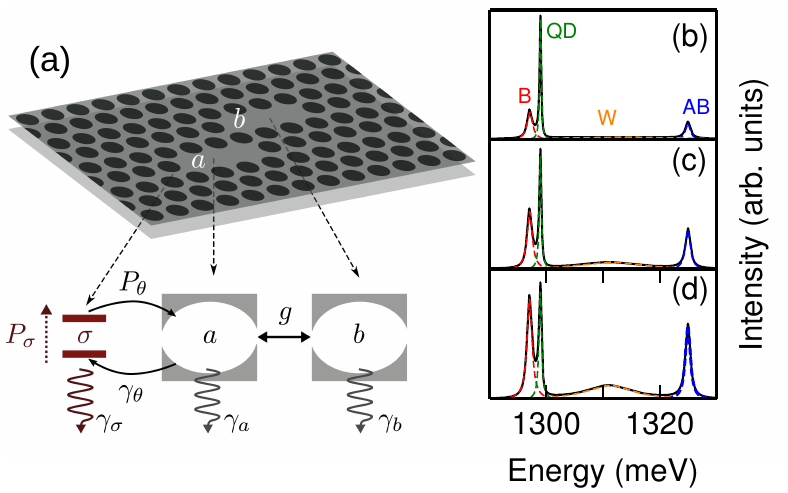}
  \caption{(a) Sketch of the model describing the system: a
    QD~$\sigma$ coupled to cavity~$a$, itself part of a photonic
    molecule with cavity~$b$. Straight solid line is the Hamiltonian
    coherent coupling, curved solid lines is the Liouvillian coherent
    coupling that comes from the phonon-mediated transitions, dotted
    line is the incoherent pumping, wavy lines are the decay
    terms. (b-d) PM emission spectrum as a function of pumping (solid
    black lines) and its decomposition in terms of the various modes
    involved (colored lines). The couple of Rabi doublets from
    Eqs.~(\ref{eq:domoct9135429CEST2016}) that can coexist gives rise
    to a triplet structure.}
  \label{fig:saboct8162443CEST2016}
\end{figure}
In our case, the key of the puzzle involves the strong, efficient and strongly
asymmetric phonon-assisted coupling mechanism already characterized in such
solid-state platforms~\cite{hohenester09a}.  The molecule finds itself in either
one these two scenarios: with the QD that excited it in the first place, through
a phonon-assisted process, now in its ground state, or, on the contrary, still
in its excited state. Both situations are possible because there is a
probabilistic aspect to both the excitation and emission of the various
components so that the QD can be re-excited before the molecule gets de-excited.
Nevertheless, this incoherent transfer of excitation is correlated, and this is
a crucial element of the model.  If the QD is de-excited, the molecule finds in
the empty QD an efficient decay channel that brings it in weak-coupling.  On the
other hand, if the QD is still excited, the probability for the molecule to
decay through this channel gets suppressed and it therefore retains its strong-
coupling.  We provide a simple theoretical model that produces this rich and
unexpected phenomenology according to the mechanism we have just described.
Unlike the model based on the multi-excitonic structure of the quantum dot
\cite{yamaguchi09a}, our mechanism holds with a simple two-level system. The
Hamiltonian itself is the simplest possible one to capture the key dynamics of
our system: two cavities~$a$ and~$b$ coupled with a strength~$g$ much larger
than the coupling~$g_\sigma$ of a QD~$\sigma$ coupled to one cavity only. The
Hamiltonian describing this situation reads
\begin{equation}
  \label{eq:vieoct7010751CEST2016}
  H = \omega_a a^\dagger a + \omega_b b^\dagger b + \omega_\sigma \sigma^\dagger \sigma + g(a^\dagger b + b^\dagger a) + g_\sigma (a^\dagger \sigma + \sigma^\dagger a)\,.
\end{equation}
The dynamics is described with a master
equation~$i\partial_t\rho=[H,\rho]+\mathcal{L}\rho$ for the total
density matrix~$\rho$, where the Liouvillian~$\mathcal{L}$ takes the
form:
\begin{equation}
\label{eq:vieoct7011429CEST2016}
\mathcal{L}\equiv\gamma_a \mathcal{L}_a + \gamma_b \mathcal{L}_b + \gamma_\sigma \mathcal{L}_\sigma + P_\sigma \mathcal{L}_\sigma + P_\theta \mathcal{L}_{\sigma a^\dagger} + \gamma_\theta \mathcal{L}_{\sigma^\dagger a}\,,
\end{equation}
where $\mathcal{L}_c\rho$ is the superoperator that is defined, for a
generic operator~$c$, as
$c^\dagger c\rho+\rho c^\dagger c-2c\rho c^\dagger$.
Equation~(\ref{eq:vieoct7011429CEST2016}) describes, respectively, the
cavities~$a$ and $b$ lifetime, the QD lifetime and rate of excitation
and, crucially, an incoherent coupling mechanism between the QD and
cavity~$a$ leading to the excitation of the cavity by the QD (at
rate~$P_\theta$) or on the opposite to its de-excitation by
transferring back the excitation to the QD (at
rate~$\gamma_\theta$). Importantly, however, this phonon-cavity
coupling is correlated as arising from a simultaneous transfer of the
excitation from the QD to the cavity, or vice-versa, as mediated by a
phonon. These terms arise for instance from the phonon-mediated
coupling studied experimentally and modelled theoretically by Majumdar
\emph{et al.}~\cite{majumdar11a} to account for this type of cavity
feeding in microcavity QED. A sketch of this model is shown in
figure~\ref{fig:saboct8162443CEST2016}(a). Note that the incoherent
version of the QD-`cavity~$a$' coupling allows different rates of
excitation transfers, unlike the Hamiltonian case where the flow back
and forth has the same rate~$g_\sigma$. This is actually one of the
important features of the model as the W peak is produced in
conditions where $\gamma_\theta\gg P_\theta$. In fact, this condition
is more important for producing a state-dependent configuration of the
molecule emission than the saturable two-level character of the QD.
The Diagonalisation of the master equation leads to two Rabi
splittings for the QD--PM system:
\begin{subequations}
  \label{eq:domoct9135429CEST2016}
  \begin{align}
    R_1&= \mathrm{Re} \sqrt{g^2- \left((\gamma_a-\gamma_b)/4\right)^2}\,,\label{eq:sáboct8112811CEST2016}\\
    R_2&= \mathrm{Re} \sqrt{g^2- \left((\gamma_a+\gamma_\theta-\gamma_b)/4\right)^2}\label{eq:sáboct8112818CEST2016}\,.
  \end{align}  
\end{subequations}
The first expression, Eq.~(\ref{eq:sáboct8112811CEST2016}), is the
standard Rabi splitting between the A and AB modes. The second
expression, Eq.~(\ref{eq:sáboct8112818CEST2016}), is similar but
absorbs the phonon decay-rate $\gamma_\theta$ into the effective decay
rate of the cavity that is coupled to the dot. The master equation
shows that both of these Rabi rates enter the dynamics.  This provides
a quadruplet structure to the emission spectrum. However, the
broadening corresponding to the weaker $R_2$-splitting is larger than
that corresponding to~$R_1$ and this makes difficult to resolve
spectrally four peaks. Instead, one obtains features of a
weakly-coupled system in the form of a broad single central peak.
This is the structure we observe in the experiment as the W peak,
although with hindsight, one could also recognize signs of a
quadruplet for instance in figures~\ref{figure4_T_tuning}(a)
and~\ref{figure5_TR_spec}(a). There, a doublet is apparent, although
it is of imbalanced height, just as, however, the outer doublet (which
could be due to a slight detuning or other variations from an ideal
light-matter coupling scenario).  At vanishing pumping, the
phonon-mediated transfer is small and so is~$\gamma_\theta$,
making~$R_1\approx R_2$ and there is only room for the expected,
conventional strong-coupling picture of a Rabi doublet, as shown in
figures~\ref{figure1_SEM}(c)~and~\ref{fig:saboct8162443CEST2016}(b).
As the pumping is increased, the new decay channel that appeared can
be so strong as to bring the molecule in the weak-coupling regime.
\begin{figure}
    \centering
    \includegraphics{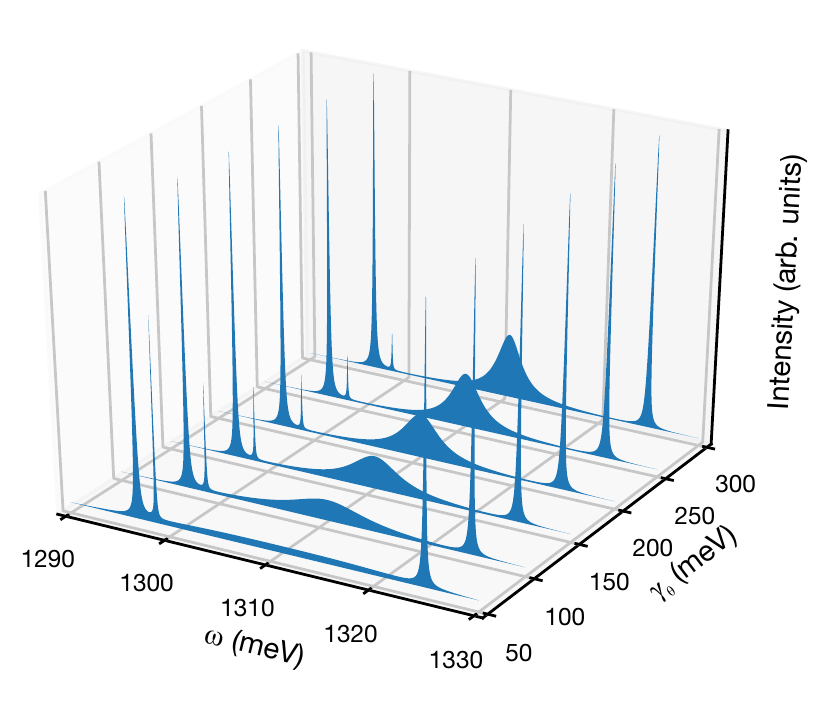}
    \caption{Calculated emission spectrum of the system as a function of
      $\gamma_\theta$, for the parameters
      $g=\unit{14}{\milli\electronvolt}$,
      $g_{JC}=\unit{0.14}{\milli\electronvolt}$,
      $\gamma_a=\gamma_b=\unit{0.23}{\milli\electronvolt}$,
      $\gamma_\sigma=\unit{0.001}{\milli\electronvolt}$,
      $E_g(0)=\unit{1299.0}{\milli\electronvolt}$ and
      $\omega_c(T=0)=\unit{1311}{\milli\electronvolt}$.}
    \label{fig:saboct8171449CEST2016}
\end{figure}
What is remarkable is that this does not destroy the Rabi doublet,
however, since this decay channel is conditional on the state of the
QD, which can saturate and stops perturbing the molecule's dynamics,
which recovers its strong-coupling. The necessity in the model of the
correlated character for the excitation transfer between the QD and
the PM as well as the existence of two Rabi splittings depending on
the state of the QD suggest an analogy with phonon-sidebands, that are
produced as the result of an optical transition affecting its
surrounding matrix. Although phonons are also responsible in our case
for making this scenario possible, the surrounding matrix itself is
actually the QD and we have therefore a 0D counterpart of this
phenomenon.

There is an excellent qualitative agreement between our simple
minimalistic model and the experimental data, with all the notable
features being reproduced. Figure~\ref{fig:saboct8171449CEST2016}, for
instance, gives an overview of the spectral features as a function of
increasing phonon-induced coupling~$\gamma_\theta$, showing the neat
transition from a conventional Rabi doublet for the PM in presence of
a sharp and dominating QD line at low pumping, to a triplet with the
added W line and a weakening contribution from the QD line at high
pumping, as is observed experimentally
(cf.~figure~\ref{figure3_non_res_PL}).
\begin{figure}
    \centering
    \includegraphics[scale=1]{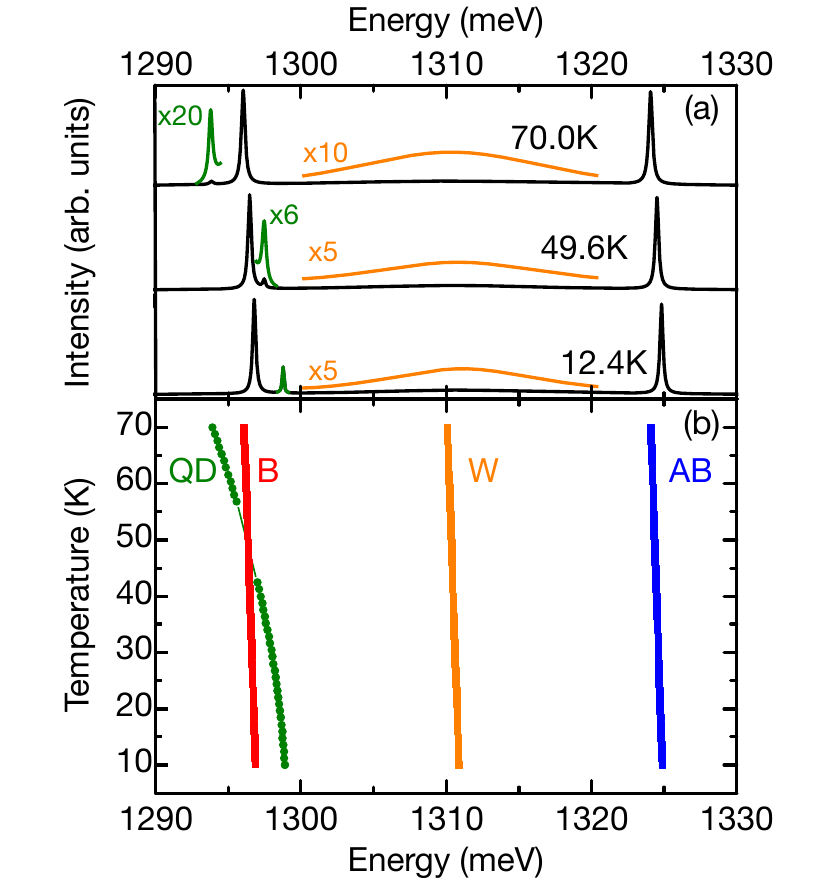}
    \caption{Emission spectrum of the photonic molecule system at
      different temperatures. Panel (a) $T=\unit{70}{\kelvin}$,
      $T=\unit{49.6}{\kelvin}$ and 
      $T=\unit{12.4}{\kelvin}$. Additionally, panel (b) shows the
      center of the peaks as a function of the temperature for the
      parameters $g=\unit{14}{\milli\electronvolt}$,
      $g_\sigma=\unit{0.14}{\milli\electronvolt}$,
      $\gamma_a=\gamma_b=\unit{0.23}{\milli\electronvolt}$,
      $\gamma_\sigma=\unit{0.001}{\milli\electronvolt}$,
      $E_g(0)=\unit{1299.0}{\milli\electronvolt}$,
      $\omega_c(T=0)=\unit{1311}{\milli\electronvolt}$ and
      $P_\sigma=\unit{0.08}{\milli\electronvolt}$.}
    \label{fig:Mon19Mar012413GMT2018}
\end{figure}
\begin{figure}[h!]
    \centering
    \includegraphics[scale=1]{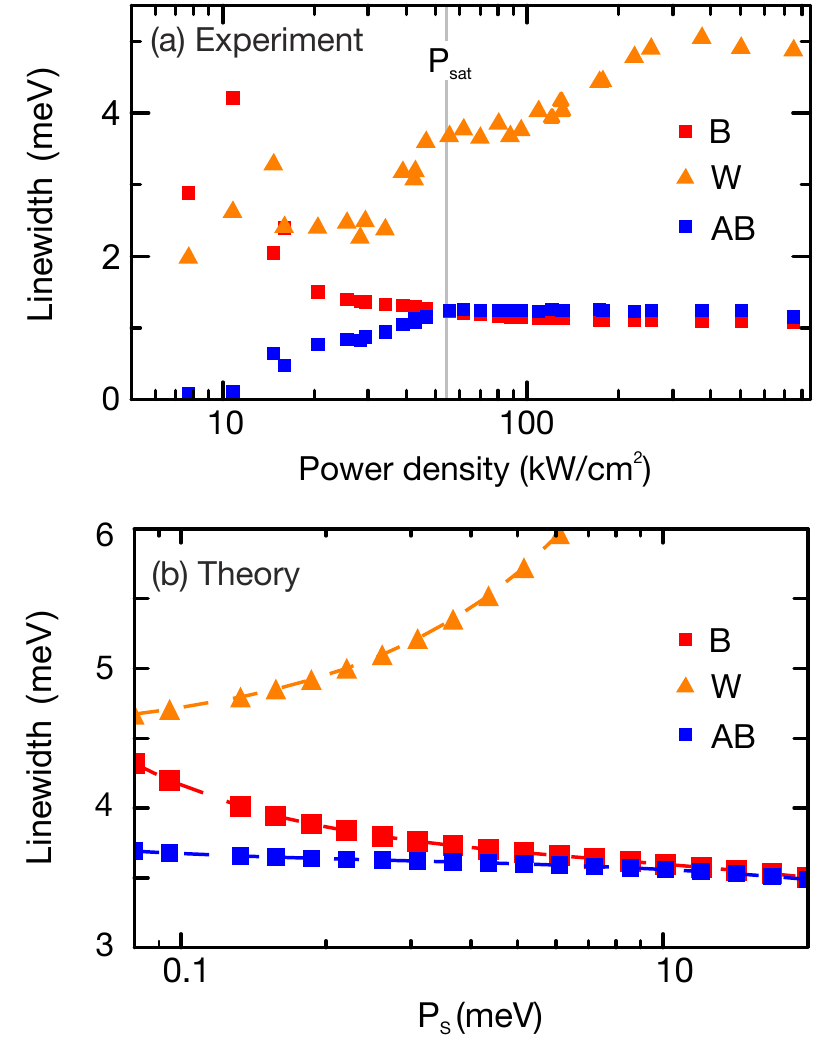}
    \caption{Linewidths as a function of the power density of the
      three emission peaks. The labels $B$ indicates bonding as red
      squares, $AB$ the anti-bonding as blue squares and $W$ central
      peak as orange triangles. The quantum dot is in resonance with
      the bonding mode.  Experimental measurements are shown in panel (a), where the grey line highlights the saturation power $P_{sat}$ of the B and AB cavity modes. The numerical calculations based on our theoretical model are shown in panel (b) where the parameters used are
      $g=\unit{14}{\milli\electronvolt}$,
      $g_{JC}=\unit{0.14}{\milli\electronvolt}$,
      $\gamma_a=\gamma_b=\unit{2.3}{\milli\electronvolt}$,
      $\gamma_\sigma=\unit{0.6}{\milli\electronvolt}$,
      $P_\theta=\unit{0.18}{\milli\electronvolt}$,
      $\gamma_\theta=\unit{360}{\milli\electronvolt}$ and
      $T=\unit{75}{\kelvin}$.}
    \label{fig:fig-linewidth}
\end{figure}
Similarly, Fig.~\ref{fig:Mon19Mar012413GMT2018} shows how the temperature
dependence matches with the experimental observation in
Fig.~\ref{figure2_mode_PLE}, and in Fig.~\ref{fig:fig-linewidth}, the linewidths dependence
are contrasted, with a good qualitative agreement.  While there can be
some quantitative differences, these are probably due to the fact that
in the actual experiment, the key variables $P_\sigma$
and~$\gamma_\theta$ are expected to be interconnected, while in the
model, they are independent free parameters that we typically vary one
at a time. In these conditions, it would be time consuming to aim for
a fit of the data, that is also not guaranteed to be excellent since
we have privileged a simple phenomenological model to capture the
physics involved, rather than a more accurate but possibly also more
confusing full semiconductor model that could provide such a
quantitative agreement. In any case, the theoretical model
unambiguously describes characteristic and distinctive features, and
those are consistently observed in the experiment. The main
consequences of this understanding of the anomalous W peak are thus
that our system allows for a coexistence of weak and strong coupling,
without either regime overtaking the other.
This gives rise to a new regime of light-matter interactions with
strong qualitative hallmarks, that have been observed thanks to the
versatility and richer environment provided by a solid-state platform.

\section{Conclusions}

We have studied the rich phenomenologies that occur in solid-state
cQED experiments involving a QD coupled to a photonic molecule. We
conducted a comprehensive experimental characterization of the
structure using several techniques and in various regimes of
excitation. At high pumping, we observed an unexpected feature in the
form of an anomalous peak~W that is energetically between the PM Rabi
doublet. This peak, that bears all the features of a cavity mode, is
explained by a simple phenomenological model of light-matter coupling
between the QD and the PM that involves a type of coupling
(phonon-mediated) that makes the molecule emit in two distinctive
environments, that allow or on the opposite impede its
strong-coupling.  This results in a coexistence of both regimes, as is
described theoretically by a simple phenomenological model that
reduces the problem to its key ingredients.  From this model, we can
identify which elements are necessary from those that do not alter the
phenomenological observation. For instance, the phonon-assisted
Liouvillian coupling terms $\mathcal{L}_{\sigma^\dagger a}$ and
$\mathcal{L}_{\sigma a^\dagger}$, describing incoherent, but
correlated, transfers of excitation, are required, as mere rate
equations do not reproduce this dynamical dependency of the molecule's
emission on the state of the QD, and only one regime is observed at a
time (in which case a triplet would be observed if each regime could
be established for long periods of time as compared to the system's
dynamics but short as compared to the integration time, as previously
discussed in the literature~\cite{hennessy07a}). Our observations show how the richer and highly tunable geometries
that are made possible by solid-state microcavity QED can give rise to
new regimes of light-matter interactions that bring curious variations
on otherwise familiar themes.

\paragraph*{Acknowledgements:} We gratefully acknowledge financial
support from the DFG via SFB-631, Teilprojekt B3 and the german
excellence initiative via the Nanosystems Initiative Munich, the BMBF
via Project No. 16KIS0110, part of the Q.com-Halbleiter consortium and
the Ministry of Science and Education of the Russian Federation
(RFMEFI61617X0085). S.E.-A. and H.V.-P. gratefully acknowledge funding
by COLCIENCIAS project ``Emisi\'on en sistemas de Qubits
Superconductores acoplados a la radiaci\'on. C\'odigo 110171249692, CT
293-2016, HERMES 31361''. S.E.-A. acknowledges support from the ``Beca
de Doctorados Nacionales de COLCIENCIAS 727''. E.A.G acknowledges
financial support from the Vicerrector\'ia de Investigaciones,
Universidad del Quind\'io.

%############################ Bibliography #####################################
\FloatBarrier
\bibliographystyle{apsrev}
%\bibliography{Sci}% Produces the bibliography via BibTeX.

%%% Local Variables:
%%% mode: latex
%%% TeX-master: t
%%% End:

\end{document}